\documentstyle[11pt]{article}
\newcommand{\IM}{{\rm Im}}

\begin{document}

\centerline{\large\bf Test of T violation in neutral B decays 
	\footnote{Supported in part by National Natural Science 
	Foundation of China } } \vspace{1cm}

\centerline{ Dongsheng Du ~~and ~~ 
             Zhengtao Wei }

\vspace{1cm}

\begin{center}
CCAST(World Laboratory), ~~P.O.Box $8730$,
Beijing $100080$, China 

and

Institute of High Energy Physics, P.O.Box $918(4)$, Beijing
$100039$, China
\end{center}

\vspace*{0.3cm}

\begin{center} \begin{minipage}{12cm}

\noindent{\bf Abstract}

T violation should be tested independently of CP violation.  
Besides K system, B meson decays provide another good place to 
study T violation. In the Standard Model,  T violation in 
$B^0 \rightleftharpoons \bar{B^0}$ oscillation is expected 
to be small.  The angular distribution of $B\to VV$ decay 
permits one to extract the T-odd correlation. In the absence 
of final state interaction, T violation in 
$B\to J/\psi(l^+ l^-) K^*(K_S\pi^0)$ 
decay can reach $4-7\%$ via $B^0-\bar{B^0}$ mixing.

\end{minipage} 
\end{center}

\vspace{1.5cm}

\newpage

\baselineskip 24pt

\section*{1.  Introduction}

Under the assumption of CPT invariance, the observed CP violation in neutral 
K decays demonstrates T violation in weak decays. However, T violation 
and CP violation are different physics concepts. CP violation requires the 
partial rate difference of the particle and its antiparticle while T violation 
needs the partial difference between the decay process and its time
reversed process. The evidence for T violation based on CP violation 
and CPT invariance is indirect. The test of T 
violation should be done independently of CP violation. 

Recently, CPLEAR collaboration gives the first direct observation of T 
violation in the difference of probability between $K^0\to \bar{K^0}$ 
and $\bar{K^0}\to K^0$ in the limit of CPT symmetry and 
the validity of $\Delta S=\Delta Q$ rule \cite{CPLEAR}. Moreover, KTeV 
observed another evidence of T violation in the planar-angle asymmetry 
in the $K_L\to \pi^+\pi^- e^+ e^-$ decay \cite{KTeV}. 
Although the validity of T violation 
in these two decays  was questioned by \cite{question}, we expect the future 
experiments 
can exclude some questions about T violation. Up to now, the study 
of T violation is mainly in K decays. It is well-known that  B meson 
decays can provide another good place for testing CP violation. 
If the CPT invariance holds, T violation should be exactly 
equal to CP violation. Since CP violation in neutral B decays can be 
large (${\cal O}(1)$), the large T violation may  happen in B decays. 
So it is necessary to study T violation in B decays. 

For the weak decay process $i\to f$, T violation is defined by 
$\Delta_T\equiv \frac{\Gamma(f'\to i')-\Gamma(i\to f)}
{\Gamma(f'\to i')+\Gamma(i\to f)}$ where $f'\to i'$ is the reversed 
process and the prime denotes the reverse of spin and momentum. In
general, it is difficult to implement the reversed weak decay. 
One reason is that it is impossible to set up the initial 
condition \cite{Sachs} such as in nuclear beta decay. 
The other reason lies in that the weak decay is so weak that it is unable 
to extract the reversed weak decay from the strong and electromagnetic 
backgrounds. The reversed reaction of $B^0\to \pi^+ \pi^-$ is such an  
example. The only exception is the neutral particle oscillation induced 
by weak interaction such as $K^0\rightleftharpoons \bar{K^0}$ 
and $B^0\rightleftharpoons \bar{B^0}$. We shall discuss  
T violation in $B^0 \rightleftharpoons \bar{B^0}$ oscillation 
without or with the assumption CPT invariance in this paper. 
For the latter case, T violation in $B^0\rightleftharpoons \bar{B^0}$  
oscillation is predicted to be small in the Standard Model (SM). 

Another way to observe T violation is to measure the T-odd correlations
in the final states of weak decays. A T-odd correlation is one that
changes sign under the reverse of all incoming and outgoing three momentum
and polarization. One classic example is the triple T-odd correlation
${\bf \sigma_n} \cdot({\bf k_p} \times {\bf k_e})$ of the nuclear beta
decay where ${\bf \sigma_n}$ is the spin of the neutron 
and ${\bf k_p}$, ${\bf k_e}$ are the three momentum of proton and 
electron. Whether the T-odd observable is considered as T violation should
be viewed with caution. There is a mimicry of T violation caused by final 
state interaction (often refers to strong interaction) even if the 
fundamental interactions is time reversal invariant. Wolfenstein calls it 
"pseudo Time Reversal Violation (pseudo TRV)". In \cite{Lola}, 
the authors prove that 
using the unitarity constraint and CPT invariance of final state interaction, 
the T-odd effect can be identified with a measurement of T violation if 
the final state interaction effects are small and negligible. 
According to \cite{Sehgal}, the correlation between the meson and lepton 
plane in $K_L\to \pi^+\pi^- e^+ e^-$ decay is T-odd and thus violates 
time reversal symmetry. The same method is used in \cite{Kruger} to study 
the angular distribution of 
$B\to K^- \pi^+ e^+ e^-$ and $B\to \pi^- \pi^+ e^+ e^-$. 
Their results show that T violation is small. 
In the differential angular distribution of the decay $B\to V_1 V_2$
where $V_{1,2}$ represent two vector mesons, the interference terms
contain T-odd correlation contribution. If $\phi$ is the angle between
the decay planes of the two vectors, the angular correlations 
$sin2\phi$ and $sin\phi$ terms are T-odd.  In this paper, we intend to 
study T violation in $B\to VV$ decays using the angular distribution 
analysis. The final state interaction effects is also taken into 
account in the T-odd observable. 

\section*{2.  T violation in $B^0\rightleftharpoons\bar{B^0}$ oscillation}

The $\Delta B=\pm 2$ weak decay via Box diagram causes the mixing 
between $B^0$ and $\bar{B^0}$. The physical states are the superposition 
of the flavor states $|B^0>$ and $|\bar{B^0}>$. The two mass eigenstates 
in neutral $B_d$ system can be generally written by 
\begin{eqnarray}
|B_1>=\frac{1}{\sqrt{|p_1^2|+|q_1|^2}}[p_1|B^0>+q_1|\bar{B^0}>] 
\nonumber\\
|B_2>=\frac{1}{\sqrt{|p_2^2|+|q_2|^2}}[p_2|B^0>-q_2|\bar{B^0}>]
\end{eqnarray}  

In the neutral K system, the mixing parameter $p_i$, $q_i$ are usually 
represented by small parameters of $\epsilon$, $\Delta$. This  
parameterization method is not suitable to apply in neutral B system 
because CP and T violation in it is predicted to be large in CKM model. 
We take the exponential parameterization as given in \cite{Sanda}. Thus the 
mixing parameters  $p_i$, $q_i$ are related by
\begin{eqnarray}
\frac{q_1}{p_1}= tg\frac{\theta}{2}e^{i\phi}, ~~~~
\frac{q_2}{p_2}=ctg\frac{\theta}{2}e^{i\phi}  
\end{eqnarray}
where $\theta$ and $\phi$ are complex phases in general. According to 
\cite{Sanda}, T violation requires $\phi\neq 0$. 

The time evolution of the initially $|B^0>$ or $|\bar{B^0}>$ after a 
proper time $t$ is 
\begin{eqnarray}
|B^0(t)>=g_+(t)|B^0>+\bar g_+(t)|\bar{B^0}>
\nonumber\\
|\bar{B^0}(t)>=g_-(t)|\bar{B^0}>+\bar g_-(t)|B^0>
\end{eqnarray}
where 
\begin{eqnarray}
g_{\pm (t)}&=&e^{-im_B t-\frac{1}{2}\Gamma_B t}
       [ch(\frac{ix-y}{2}\Gamma_Bt)\pm 
        cos\theta sh(\frac{ix-y}{2}\Gamma_Bt)]
\nonumber\\
\bar{g}_{\pm}(t)&=&e^{-im_B t-\frac{1}{2}\Gamma_B t}
  sin\theta e^{\pm i\phi} sh(\frac{ix-y}{2}\Gamma_Bt)
\end{eqnarray}
with 
$x\equiv\frac{\Delta m_B}{\Gamma_B}, ~~~
y\equiv\frac{\Delta \Gamma_B}{2\Gamma_B}. $

The T violation in $B^0\rightleftharpoons\bar{B^0}$ oscillation is defined as 
\begin{equation}
A_T(t)\equiv \frac{P_{B^0(t)\to \bar{B^0}}-P_{\bar{B^0}(t)\to B^0}}
  {P_{B^0(t)\to \bar{B^0}}+P_{\bar{B^0}(t)\to B^0}}
  =\frac{|e^{i\phi}|^2-|e^{-i\phi}|^2}
   {|e^{i\phi}|^2+|e^{-i\phi}|^2}
  =-2\IM\phi
\end{equation}

From Eq.(5), $A_T(t)$ is independent of $t$. It is a constant number. 
This is the same as the case in the CPLEAR experiment. Moreover $A_T(t)$
is not related with the CPT violation parameter $\theta$. Note that 
$A_T(t)$ is proportional to $Im\phi$, the imaginary part of the mixing 
parameter $\phi$. As it will be seen later, T violation in the interference 
of the $B^0-\bar{B^0}$ mixing and the decay amplitude  requires 
$Re\phi \neq 0$.   

The experimental test of T violation $A_T(t)$ can be determined in the 
semileptonic decay and the same-sign dileptonic ratios of B decays through 
\cite{Wei}
\begin{eqnarray}
A_{T}(t)=\frac{\Gamma(B^0(t)\to \bar Xl^-\nu)-
   \Gamma(\bar{B^0}(t)\to Xl^+\nu)}
  {\Gamma(B^0(t)\to \bar Xl^-\nu)+
   \Gamma(\bar{B^0}(t)\to Xl^+\nu)}
  =\frac{N^{++}-N^{--}}{N^{++}+N^{--}}
  =-2\IM\phi
\end{eqnarray}
where $N^{++}$, $N^{--}$ are the same-sign dilepton events. In Eq.(6), 
The validity of $\Delta B=\Delta Q$ rule is assumed. 

We next turn to discuss the SM expectation of the $A_T$. In SM, CPT 
is invariant, and the origin of CP and T violation lies in the nonzero 
complex phase in CKM matrix. The mixing parameter $\theta$ in Eq.(2) 
will be equal to $\frac{\pi}{2}$.  
Thus, $\frac{q_1}{p_1}=\frac{q_2}{p_2}=\frac{q}{p}$. 
According to \cite{Nir}, 
\begin{eqnarray}
|\frac{q}{p}|-1=|e^{i\phi}|-1=\frac{1}{2}
Im\frac{\Gamma_{12}}{M_{12}}\sim {\cal O}(10^{-3})
\end{eqnarray}
Thus
\begin{eqnarray}
A_T(t)\approx Im\frac{\Gamma_{12}}{M_{12}}\sim {\cal O}(10^{-3})
\end{eqnarray}

The above estimate is based on the assumption that the box diagram with a
cut is appropriate to calculate $\Gamma_{12}$.  The uncertainty  from 
the use of quark diagram to describe $\Gamma_{12}$ could be a factor of 
2-3.

\section*{3. T violation in the angular distribution of $B\to VV$ decay}

As discussed in the Introduction, another way to observe T violation is
through the T-odd correlation in the final states. 
Nonleptonic B decays play important role in exploring CP violation such 
as the decays $B\to J/\psi K_S$, $\pi\pi$, $\pi K$, etc. 
Unlike CP violation,  
there is no T-odd correlation in $B\to PP$ and $B\to VP$ decay. Because 
the decay amplitude contains only T-even term: 
the momentum square for $B\to PP$ decay; and the product of 
momentum and polarization vector for $B\to VP$ decay. 
Both of these terms are invariant under time reversal.  
In $B\to VV$ decays, the angular correlation between the decay planes of
two vectors contains T-odd terms  thus provides place to search T
violation. We take $B\to J/\psi K^*$ as an 
example to discuss the T violation in $B\to VV$ decays. 

The differential decay distribution for $B\to K^* J/\psi\to (K\pi)(l^+ l^-)$ 
is \cite{Kramer}:
 \begin{eqnarray}
\frac{d^3 \Gamma}{dcos\theta_1 dcos\theta_2 d\phi}
  &=& \frac{|\stackrel{\rightarrow}{p}|}{16\pi^2 m_B^2}\frac{9}{8}
    \{\frac{1}{4}sin^2\theta_1(1+cos^2\theta_2)(|H_{+1}|^2+|H_{-1}|^2)
     +cos^2\theta_1 sin^2\theta_2 |H_0|^2
\nonumber\\
  & & -\frac{1}{2}sin^2\theta_1 sin^2\theta_2 [cos2\phi Re(H_{+1}H^*_{-1})
      -sin2\phi Im(H_{+1}H^*_{-1}) ] \\
  & & -\frac{1}{4}sin2\theta_1 sin2\theta_2 [cos\phi Re(H_{+1}H^*_0
     +H_{-1}H^*_0)-sin\phi Im(H_{+1}H^*_0-H_{-1}H^*_0) ] \} 
\nonumber
\end{eqnarray}
where $\theta_1$ is the polar angle of the $K$ momentum in the rest frame of
the $K^*$ meson with respect to the helicity axis of $K^*$ meson   
(the negative of the the direction of the $J/\psi$ in $K^*$ rest frame) 
and similarly $\theta_2$ is the polar angle of the positive lepton $l^+$
($e^+$ or $\mu^+$) momentum in the rest frame of the $J/\psi$  with respect
to the helicity axis of $J/\psi$; $\phi$ is the angle between
the planes of the two decays of $K^*\to K\pi$ and $J/\psi\to l^+ l^-$. 
In eq.(9), $\stackrel{\rightarrow}{p}$ is the three momentum of the vector
$K^*$; $H_i$ are the helicity amplitude defined in \cite{Kramer}.
The angle correlations $sin2\phi$ and $sin\phi$ are T-odd. 
To confirm this, define the unit vector $\hat{\bf p}\equiv
\frac{\stackrel{\rightarrow}{p_{K^*}}}
{|\stackrel{\rightarrow}{p_{K^*}}|}$. Thus, 
\begin{eqnarray}
sin\phi&=& ~~
(\frac{\stackrel{\rightarrow}{p_{K}}
         \times\stackrel{\rightarrow}{p_{\pi}} }
{|\stackrel{\rightarrow}{p_{K}}
         \times\stackrel{\rightarrow}{p_{\pi}} |} )\times
(\frac{\stackrel{\rightarrow}{p_{l^+}}
         \times\stackrel{\rightarrow}{p_{l^-}} }
{|\stackrel{\rightarrow}{p_{l^+}}  
         \times\stackrel{\rightarrow}{p_{l^-}} |} )
\cdot \hat{\bf p}
\nonumber \\
sin2\phi&=& 2
(\frac{\stackrel{\rightarrow}{p_{K}}
         \times\stackrel{\rightarrow}{p_{\pi}} }
{|\stackrel{\rightarrow}{p_{K}}
         \times\stackrel{\rightarrow}{p_{\pi}} |} )\times
(\frac{\stackrel{\rightarrow}{p_{l^+}}
         \times\stackrel{\rightarrow}{p_{l^-}} }
{|\stackrel{\rightarrow}{p_{l^+}}
         \times\stackrel{\rightarrow}{p_{l^-}} |} )
\cdot \hat{\bf p}
(\frac{\stackrel{\rightarrow}{p_{K}}
         \times\stackrel{\rightarrow}{p_{\pi}} }
{|\stackrel{\rightarrow}{p_{K}}
         \times\stackrel{\rightarrow}{p_{\pi}} |} )\cdot
(\frac{\stackrel{\rightarrow}{p_{l^+}}
         \times\stackrel{\rightarrow}{p_{l^-}} }
{|\stackrel{\rightarrow}{p_{l^+}}
         \times\stackrel{\rightarrow}{p_{l^-}} |} )
\end{eqnarray}
From the above equation, $sin2\phi$ and $sin\phi$ contain 9 and 5 
momentum vectors in the products respectively. 
Under the time reversal transformation, they change their signs. 
All these momentums are defined in the rest frame of $B$ meson. 

Another form of the angular distribution based on the transversity 
variable is given in \cite{Dunietz} \cite{Fleischer}. In that form, 
the CP-even and odd and T-even and old component is obvious. Both the 
two froms 
are principally the same except for the adoption of different variable. 
  
The integration over angles $\theta_1$ and $\theta_2$ yields the $\phi$ angle 
distribution 
\begin{eqnarray}
\frac{d\Gamma}{d\phi}=\frac{|\stackrel{\rightarrow}{p}|}{16\pi^2 m_B^2}
  \{|H_{+1}|^2+|H_{-1}|^2+|H_0|^2-cos2\phi Re(H_{+1}H^*_{-1})
      +sin2\phi Im(H_{+1}H^*_{-1}) \}
\end{eqnarray}
From Eq.(11), only one T-odd $sin2\phi$ term  is left when integrating 
over angles $\theta_1$ and $\theta_2$. The other T-odd $sin\phi$ term 
can be extracted from the full three-angle distribution or the 
difference of $\phi$ angle distribution between the same hemisphere 
events (e.g. $0<\theta_1, \theta_2<\frac{\pi}{2}$) 
and the opposite hemisphere events (e.g. $0<\theta_1<\frac{\pi}{2}, ~~ 
\frac{\pi}{2}<\theta_2<\pi$). For this case, the full angle distribution 
is required to be known from the experiment. In this paper, we restrict 
our discussion in the single angle $\phi$ distribution given by Eq.(11) 
because it is easier to treat in the experiment.

So, T violation is given by 
\begin{eqnarray}
\Delta_T=\frac{( \int_{0}^{\frac{\pi}{2}} -\int_{\frac{\pi}{2}}^{\pi}
  +\int_{\pi}^{\frac{3\pi}{2}} -\int_{\frac{3\pi}{2}}^{2\pi}) 
  \frac{d\Gamma}{d\phi}d\phi }
 {( \int_{0}^{\frac{\pi}{2}} +\int_{\frac{\pi}{2}}^{\pi}
  +\int_{\pi}^{\frac{3\pi}{2}} +\int_{\frac{3\pi}{2}}^{2\pi} ) 
  \frac{d\Gamma}{d\phi}d\phi }
  =\frac{2}{\pi}\frac{Im(H_{+1}H^*_{-1})}{|H_{+1}|^2+|H_{-1}|^2+|H_0|^2}
  \equiv \frac{2}{\pi}\beta_2
\end{eqnarray}

Up to now, our analysis is model independent. In the remainder of the
paper, we will restrict our discussion in the Standard Model. From
Eq.(12), T violation observable $\Delta _T$ is proportional to the
angular correlation coefficient $\beta_2$. This relation is a general
result of the decay $B\to VV$. The nonvanishing $Im(H_{+} H_{-1}^*)$
is caused by weak CKM phases  or strong final state interaction
phases  under the condition that they contribute differently to
$H_{+1}$ and $H_{-1}$. First, we discuss the case that the final state
interaction is absent. In \cite{Kramer}, the authors had systematically 
calculated 
all the $B\to VV$ decays. Their result shows that $\beta_2$ is very
small. For most $B\to VV$ process, $\beta_2$ is less than $10^{-4}$.
In the special example of $B\to K^* J/\psi$, the tree and the
dominant QCD Penguin diagram have the same CKM phase, thus $\beta_2$ is 
nearly zero. 

Second, we consider the nonvanishing $\beta_2$ caused only by strong 
final state interaction. Since the strong interaction is T invariant, 
the violation induced by final state interaction is not the true T violation
but the mimicry of T violation.  

Let us introduce 
\begin{eqnarray}
H_{||}=\frac{1}{\sqrt 2}(H_{+1}+H_{-1})
\nonumber \\ 
H_{\bot}=\frac{1}{\sqrt 2}(H_{+1}-H_{-1})
\end{eqnarray}
and define the strong phase difference  $\delta\equiv Arg(H_{||}H^*_{\bot})$
then 
\begin{eqnarray}
\Delta_T=\frac{2}{\pi}\frac{|H_{||}||H_{\bot}|sin\delta}
   {|H_{||}|^2+|H_{\bot}|^2+|H_0|^2}
\end{eqnarray}

The fact that the T violation mimicry appears to be proportional to
$sin\delta$ was 
revealed long time ago (see \cite{Sachs}). Here we again find this 
particular 
characteristic in $B\to VV$ decays. Up to now, definite quantitative 
analysis of final state interaction has not been accessible yet. 
The concrete information about strong phase is unknown. But, based on 
some phenomenological consideration, the final state interaction effects 
in $B\to K^* J/\psi$ is estimated to be small. The small width of 
$J/\psi$ makes the strong coupling between $J/\psi$ and strong states 
be small. Moreover, there are few channels with large branching ratios 
that can transform into the final state $K^* J/\psi$ through strong 
interaction. From above, it seems that T violation in the angular
distribution of $B\to K^* J/\psi$ is  a small effect. 

There are three types of CP violation in neutral B system. CP violations 
in decay amplitude and mixing are small. The most important type is the 
CP violation in the interference of mixing and decay. This type 
is promising to give large CP violation. T violation in 
$B^0\rightleftharpoons\bar{B^0}$ 
oscillation  is due to $B^0-\bar{B^0}$ mixing. T violation in the above 
discussion of the $B\to VV$ decay which requires weak phases contribute 
differently to helicity amplitudes belongs to the T violation in decay. 
Both of them are small. Next, we intend to seek for large T 
violation in the interference of mixing and decay. 

In the Standard Model, the time evolution of $B^0$ and $\bar{B^0}$ is 
obtained from Eq.(3) 
\begin{eqnarray}
|B^0(t)>=f_+(t)|B^0>+\frac{q}{p} f_+(t)|\bar{B^0}>
\nonumber\\
|\bar{B^0}(t)>=\frac{p}{q}f_-(t)|B^0>+f_+(t)|\bar{B^0}>
\end{eqnarray}
where
$f_+(t)=e^{-im_B t-\frac{\Gamma_B t}{2}}cos\frac{\Delta m_B t}{2}$, ~~~
$f_-(t)=e^{-im_B t-\frac{\Gamma_B t}{2}}isin\frac{\Delta m_B t}{2}$, ~~~
$\frac{q}{p}=e^{-i2\beta}$, and $\beta$ is the angle of the unitarity triangle. 
We have neglected the width difference of  two mass eigenstates. 

If $K^{*0}$ in the decay $B^0\to K^{*0}J/\psi$ is observed to 
decay to CP eigenstates $\pi^0 K_S$, then angular distribution analysis 
gives the time dependent T violation as 
\begin{eqnarray}
\Delta_T(t)&=&\frac{2}{\pi}\frac{Im(H_{||}(t)H_{\bot}^*(t))}
   {|H_{||}(t)|^2+|H_{\bot}(t)|^2+|H_0(t)|^2} \\
&  =&\frac{2}{\pi}\frac{ |H_{||}(0)||H_{\bot}(0)|
   [sin\delta cos\Delta m_Bt+cos\delta cos2\beta sin\Delta m_Bt]
   e^{-\Gamma_B t} }
   {[|H_{||}(0)|^2+|H_{\bot}(0)|^2+|H_0|^2(0)
   +sin2\beta sin\Delta m_Bt(|H_{||}(0)|^2+|H_{0}(0)|^2-|H_{\bot}(0)|^2)]
   e^{-\Gamma_B t} } \nonumber
\end{eqnarray}
The above time dependent T violation has two contributions. The first term
is 
the mimicry of T violation induced by final state interaction. The second 
term which  contains $cos2\beta$ in the absence of final state interaction
is due to the interference of mixing and decay. In this case, the 
small final state interaction effects can be neglected. 

The time integrated T violation obtained from Eq.(16) is 
\begin{eqnarray}
D_T&=&\frac{2}{\pi}\frac
   {\int_{0}^{\infty}dt~Im(H_{||}(t)H_{\bot}^*(t))}
   {\int_{0}^{\infty}dt~[|H_{||}(t)|^2+|H_{\bot}(t)|^2+|H_0(t)|^2]} \\
&  =&\frac{2}{\pi}\frac{ |H_{||}(0)||H_{\bot}(0)|
   [\frac{sin\delta}{1+x^2}+cos\delta cos2\beta \frac{x}{1+x^2}]}
   {|H_{||}(0)|^2+|H_{\bot}(0)|^2+|H_0|^2(0)
   +sin2\beta\frac{x}{1+x^2}(|H_{||}(0)|^2+|H_{0}(0)|^2-|H_{\bot}(0)|^2)}
   \nonumber
\end{eqnarray}
where $x\equiv \frac{\Delta m_B}{\Gamma_B}=0.7 $ is obtained from PDG98 
\cite{PDG98}. 

In order to estimate the time integrated T violation, we use $sin2\beta=0.5$ 
and the parameters given in \cite{Fleischer} which based on the models of BSW, 
Soares and Cheng \cite{BSW}. Table 1 gives the results of time integrated 
T violation in $B\to K^*(\pi^0 K_S) J/\psi(l^+ l^-)$ with different 
models in the absence of final state interaction. 
From Table 1, one can see that different models give the T violation range 
from 0.04 to 0.07.

\begin{table}[hbt]
\begin{center}
Table 1. Time-integrated T violation in 
$B\to J/\psi(l^+ l^-) K^*(\pi^0 K_S)$
with different models 

\vspace{0.4cm}
\begin{tabular}{|c|c|c|c|} \hline \hline
       & BSW  &  Soares  &  Cheng  \\ \hline
  
$D_T$ & 0.038 & 0.069 & 0.047 \\ \hline
 
\end{tabular}
\end{center}
\end{table}

\section*{4. Conclusion }

In this paper, we present a study of T violation in 
$B^0 \rightleftharpoons\bar{B^0}$ 
oscillation and $B\to VV$ decays. T violation in B decays opens another 
way to test Standard Model and the origin of CP/T violation. 
In $B^0\rightleftharpoons \bar{B^0}$ oscillation, T violation induced 
by $B^0-\bar{B^0}$
mixing is about the order of $10^{-3}$. This tiny effects is possible 
to observe in semileptonic decay and dileptonic decay at B-factory and 
LHC-B. If  a large effect is found, it will be new physics beyond the
Standard Model. 
T violation in decay of $B\to VV$ from the interference term $\beta_2$ 
with different weak phases that contribute to helicity amplitudes 
is small, and this effect can not be extracted from  
the mimicry induced by final state interaction. 
Via interference of $B^0-\bar{B^0}$ mixing and decay, the time integrated
T violation in 
$B^0\to K^{*0}J/\psi \to (K_S \pi^0)(l^+ l^-)$ decay can reach 4-7\% 
which is experimentally accessible.  In this case final state interaction 
effects can be neglected. 

From the CPT theorem, T violation should be exactly equal to CP violation.
CP violation in neutral B decays can be 
as large as ${\cal O}(1)$, while T violation we had found is at the 
10\%  level of CP violation. So the question that how and where to find 
large T violation in B decays arises. 

\vspace{0.5cm}
{\it Note added:} 
After finishing this paper, we became aware that 
the T-odd correlation in $B\to VV$ decays was pointed out 
in hep-ph/9911338 \cite{VV}. However, the physics motivation 
of two papers are different.

\section*{Acknowledgment}

This work is supported in part by National Natural Science 
Foundation of China and the Grant of State Commission of Science
and Technology of China.

\end{document}